\documentclass{proc}

\usepackage[preprint]{procjac}

\usepackage{color}

\docident{\hspace{\fill}
\textit{Presented at the 2003 Particle Accelerator Conference}
\hspace{\fill}\makebox[0pt][r]{MSUCL-1275}}

\usepackage{graphics}

\newcommand{\incgraphLANDHALF}[2]{%
\makebox[\columnwidth]{%
\resizebox{#2\columnwidth}{!}{%
\includegraphics[0.35in,0.35in][5.5in,4.25in]{#1}}%
}}

\newcommand{\degree}{\ensuremath{^\circ}}

\newcommand{\etal}{{\em et al.}}

\begin{document}

\title{STATUS REPORT ON MULTI-CELL SUPERCONDUCTING CAVITY DEVELOPMENT FOR
MEDIUM-VELOCITY BEAMS\thanks{Work supported by the U.S. Department of Energy
under Grant DE-FG02-00ER41144.}}

\author{%
W.~Hartung, C.~C.~Compton, T.~L.~Grimm, R.~C.~York\\
National Superconducting Cyclotron Lab, Michigan State University, East Lansing,
Michigan, USA\\
\and
G.~Ciovati, P.~Kneisel\\
Thomas Jefferson National Accelerator Facility, Newport News, Virginia, USA}

\maketitle

%

\section{INTRODUCTION}

The Rare Isotope Accelerator (RIA) is being designed to supply an intense beam
of exotic isotopes for nuclear physics research \cite{R_JP}.  Superconducting
cavities are to be used to accelerate the CW beam of heavy ions to 400 MeV per
nucleon, with a beam power of up to 400 kW\@.  Because of the varying beam
velocity, several types of superconducting structures are needed \cite{R_IN}.

Since the RIA driver linac will accelerate heavy ions over the same velocity
range as the Spallation Neutron Source (SNS) proton linac, the 6-cell
axisymmetric 805 MHz cavities and cryostats of SNS can be used for part of the
RIA linac.  Prototypes for both SNS cavities ($\beta_g = 0.61$ and $\beta_g =
0.81$) have been tested \cite{R_JO}.  (Herein, $\beta$ is the particle velocity
divided by $c$ and $\beta_g$ is the geometric $\beta$.)

The SNS cavity design is being extended to lower velocity ($\beta_g = 0.47$)
for RIA \cite{R_JEP,R_IIP}.  Other single-cell cavities for $\beta = 0.47$ to
0.5 have also been prototyped at various laboratories \cite{R_IG,R_IH,R_ILP};
in all cases, gradients and $Q$'s have exceeded the design goals.  A 5-cell
$\beta = 0.5$ cavity has also been prototyped at JAERI \cite{R_JR}.

This paper covers the fabrication of three prototype RIA 6-cell $\beta_g =
0.47$ cavities and the RF tests on the first and second of these cavities.

\section{CAVITY DESIGN}

The SNS $\beta_g = 0.81$ and $\beta_g = 0.61$ cavities are the basis for the
RIA $\beta_g = 0.47$ cell shape \cite{R_JEP,R_IIP}.  The beam tube is enlarged
on one side of the SNS cavities in order to provide stronger input coupling. 
Less coupling is needed for RIA, so no enlargement of the beam tube is needed
for the $\beta_g = 0.47$ cavity \cite{R_JL}.  This simplifies the cavity
fabrication and yields a slight improvement in the RF parameters.  Selected
cavity parameters are given in Table~\ref{T_RF}.  In Table~\ref{T_RF}, $E_p$
and $B_p$ are the peak surface electric and magnetic field, respectively, and
$E_a$ is the accelerating gradient (transit time included) for a particle
travelling at the design velocity.

An analysis was done of the excitation of higher-order modes (HOMs) in the
cavity by the beam and coupling of the HOMs to the input coupler and pick-up
antenna.  This analysis indicates that HOM couplers are not required for
operation of the $\beta_g = 0.47$ cavity in RIA, allowing for further
simplification of the system \cite{R_JL}.

\begin{table}

\caption{Parameters of the symmetric 6-cell $\beta_g = 0.47$ cavity; $R_s$ is
the shunt impedance (linac definition).  RF quantities were calculated with
SUPERFISH \protect\cite{R_DP}. \label{T_RF}}

\vspace*{1.5ex}

\begin{center}
\begin{tabular}{|l|c|}
\hline
Mode & $TM_{010} \pi$ \\ \hline
Resonant frequency $f$ & 805 MHz \\ \hline
Cell-to-cell coupling & 1.5\% \\ \hline
$E_p/E_a$ & 3.34 \\ \hline
$c B_p/E_a$ & 1.98 \\ \hline
$R_s/Q$ & 173 $\Omega$ \\ \hline
Geometry factor & 155 $\Omega$ \\ \hline
\hline
Active length & 527 mm \\ \hline
Inner diameter at iris (aperture) & 77.2 mm \\ \hline
Inner diameter at equator & 329 mm \\ \hline
\end{tabular}
\end{center}

\end{table}

\section{SINGLE-CELL CAVITY PROTOTYPING}

Two single-cell prototypes of the $\beta_g = 0.47$ cavity were fabricated and
tested.  The highest gradient reached in the first round of tests \cite{R_JEP}
was about 15 MV/m.  The $Q$ values at 15 MV/m were about $10^{10}$; the
low-field $Q$ values were between $2 \cdot 10^{10}$ and $4 \cdot 10^{10}$. 
These measurements were done at 2 K in a vertical cryostat at Jefferson Lab.

Additional tests were done on the second of the two single-cell cavities while
commissioning the facilities at NSCL for etching, high-pressure rinsing, clean
assembly, RF testing, and helium processing of superconducting cavities.  The
highest gradient reached in these tests was about 18 MV/m, albeit with a
slightly lower $Q$; however, the $Q$ still exceeded $10^{10}$ at the design
gradient of 8 MV/m \cite{R_JMP}.

\section{MULTI-CELL CAVITY PROTOTYPING}

\subsection{Cavity Fabrication and Preparation}

Sheet Nb 4 mm in thickness with a nominal Residual Resistivity Ratio (RRR) of
250 was used for the 6-cell cavities.  The forming and joining of half-cells
were done by the standard deep drawing and electron beam welding techniques
used for SNS cavity fabrication.  As with the SNS cavities, Nb-Ti flanges and
Al alloy gaskets were used for the vacuum seal on the beam tubes.

\begin{figure}[tb]
\makebox[\columnwidth]{%
\setlength{\unitlength}{0.01in}
\begin{picture}(139,325)
\put(0,0){\rotatebox{90}{\resizebox{3.25in}{!}%
{\includegraphics{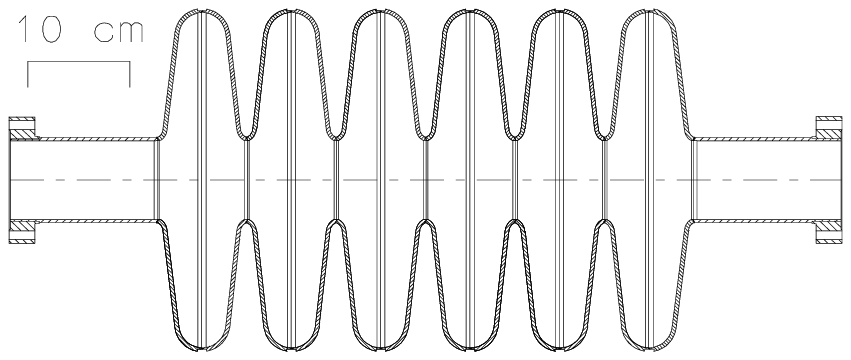}}}}
\put(0,300){\makebox(30,25){\Large(a)}}
\end{picture}
\setlength{\unitlength}{0.01in}
\begin{picture}(160,324)
\put(0,0){\includegraphics{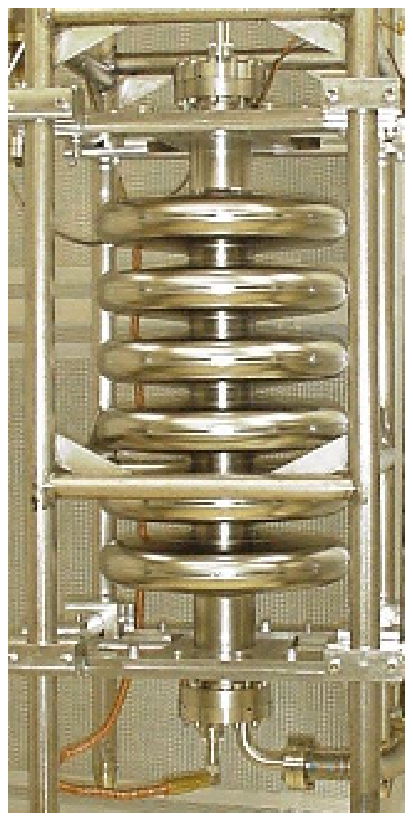}}
\put(0,300){\colorbox{white}{\makebox(30,25){\Large(b)}}}
\end{picture}
}
\vspace*{0.5ex}
\makebox[\columnwidth]{%
\setlength{\unitlength}{0.01in}
\begin{picture}(325,140)
\put(0,0){\resizebox{3.25in}{!}{\includegraphics{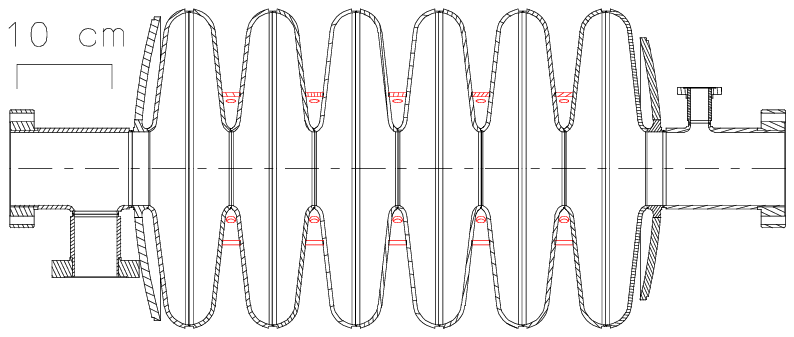}}}
\put(278,15){\makebox(30,25){\Large(c)}}
\end{picture}
}
\vspace*{0.5ex}
\makebox[\columnwidth]{%
\setlength{\unitlength}{0.01in}
\begin{picture}(306,139)
\put(0,0){\includegraphics{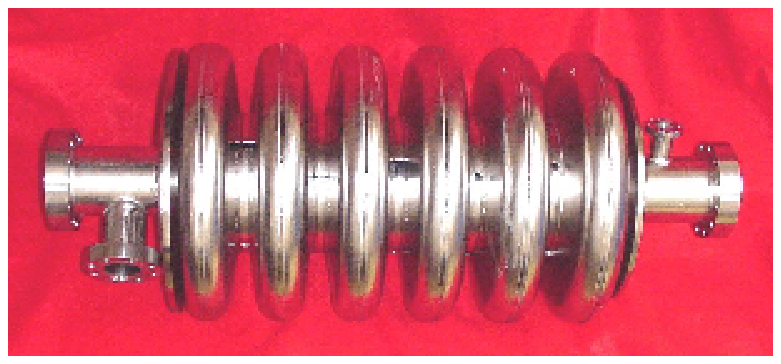}}
\put(0,114){\colorbox{white}{\makebox(30,25){\Large(d)}}}
\end{picture}
}
\caption{(a) Drawing of the first six-cell $\beta_g = 0.47$ Nb cavity and (b)
photograph of the cavity on the RF test stand.  (c) Drawing and (d) photograph
of the second cavity.}
\label{F_PICT}
\end{figure}

The first 6-cell cavity (Figures \ref{F_PICT}a and \ref{F_PICT}b) was a
simplified version without stiffening rings, dishes for attachment of the
helium vessel, or side ports for the RF couplers; these features were included
in the second and third cavities (Figures \ref{F_PICT}c and \ref{F_PICT}d).

The first cavity was etched with a Buffered Chemical Polishing  solution
to remove about 100 $\mu$m from the inside surface.  The cavity was then fired
in a vacuum furnace for 10 hours at 600\degree C to inoculate it against the
``$Q$ disease.''  The pressure in the furnace was $\leq 10^{-6}$ torr during
the heat treatment.  Field flatness tuning was done next (see below).  The
final preparation steps were etching of an additional 60 $\mu$m from the inner
surface and high-pressure rinsing with ultra-pure water in a clean room to
remove particulates from the inside surface of the cavity.

The second cavity was etched to remove 150 $\mu$m and rinsed with the
high-pressure water; it was not fired.

\subsection{Tuning}

\begin{figure}[b]
\makebox[\columnwidth]{\incgraphLANDHALF{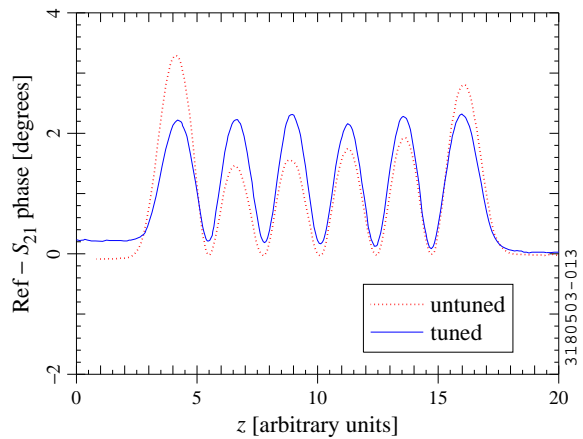}{1.0}}
\caption{Bead pulls for the second six-cell niobium cavity.}
\label{F_BEAD}
\end{figure}

Field flatness tuning was done on the first two niobium 6-cells; ancillary
tuning was also done on a 5-cell copper model.  The goal was a field unflatness
parameter ($\Delta E/E$) of 10\% or less.  The first cavity and the copper
model were tuned with a tuning jig designed for the SNS cavities.  After
tuning, $\Delta E/E$ was 7\% for the Cu cavity and 12\% for the Nb cavity.  The
second Nb cavity was tuned with a new custom-built jig for the $\beta_g =0.47$
cavity.  This made the tuning easier; a $\Delta E/E$ of 5\% was reached in one
iteration (see Figure~\ref{F_BEAD}).

\subsection{First RF Test on the First Cavity}

A vertical RF test was done on the first 6-cell cavity in September 2002.  The
cryostat was cooled down rapidly to 4.2 K and then pumped to 2 K\@.  As shown
in Figure~\ref{F_RFTEST} (squares), the low-field $Q$ was about $2 \cdot
10^{10}$ and the $Q$ remained above $10^{10}$ up to $E_a = 11$ MV/m
approximately.  A gradient of about 16 MV/m was reached.  The test was stopped
at that field due to the failure of an RF cable.  Some x-rays were observed at
high field, indicating that the decrease in the $Q$ at high field was likely
due to field emission.  Modest RF conditioning was required in order to reach a
gradient of 16 MV/m.  A small leak into the cavity vacuum manifested itself
when the cryostat was cooled down; the pressure in the cavity was about
$10^{-6}$ torr at 2 K.

\subsection{Follow-Up RF Tests on the First Cavity}

The failed RF cable was replaced, the leak in the cavity vacuum was fixed, and
the cavity was retested 1 week after the first RF test (without exposure of the
inside of the cavity to air).  A gradient of about 7 MV/m was reached.  It was
thought that helium processing might be beneficial, but the test had to be
stopped early due to scheduled maintenance of the cavity testing facility.

The next opportunity for an RF test was in January 2003.  In between tests, the
cavity was etched again to remove another 50 $\mu$m from the inner surface and
the high-pressure water rinsing was repeated.  The final filter on the
high-pressure rinsing system (between the pump and the nozzle) was temporarily
unavailable at the time of this rinse.

The results of the January 2003 test are shown in Figure~\ref{F_RFTEST}
(circles).  The low-field $Q$ was smaller than in the first test, although the
difference is within the margin of reproducibility for the measurement.  A
gradient of about 11 MV/m was reached.  The decrease in $Q$ between 9 and 11
MV/m is likely due to field emission; the x-ray signals were larger than those
seen in the first RF test.  Thus the difference between the September 2002 and
January 2003 tests could be due to particulate contamination during the
high-pressure rinse without the final filter.  Although the field level was not
as high as in the first test, the design goal of 8 MV/m was nevertheless
reached with a $Q$ in excess of $10^{10}$.

In the January 2003 tests, measurements of $Q$ as a function of gradient were
done at several different temperatures.  The low-field $Q$ at 1.8 K was higher
($2 \cdot 10^{10}$) than at 2 K, which indicates that the BCS losses are still
contributing to the surface resistance.  However, the maximum gradient at 1.8 K
was only slightly higher than at 2 K\@.  A value of $Q = 2 \cdot 10^{10}$
corresponds to a surface resistance of 8 n$\Omega$.

\subsection{RF Tests on the Second Cavity}

Vertical RF testing on the second cavity was done in May 2003.  In the first RF
test, a gradient of 8 MV/m was reached at 2 K\@.  The $Q$ was a bit low ($8
\cdot 10^{9}$) and some $Q$-switching was seen, indicating that more etching
was needed.

\begin{figure}[tb]
\makebox[\columnwidth]{\incgraphLANDHALF{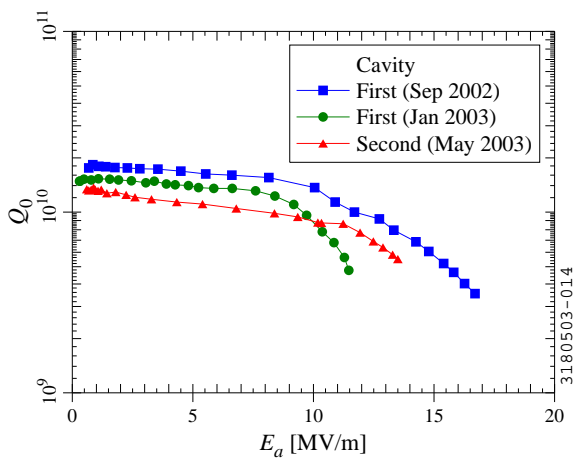}{1.0}}
\caption{Measured dependence of the quality factor on the accelerating
gradient at 2 K for the first and second 6-cell cavities.}
\label{F_RFTEST}
\end{figure}

In preparation for a second RF test, another 150 $\mu$m was removed from the
inner surface, and the high-pressure rinse was repeated.  Results from the
second RF test are shown in Figure~\ref{F_RFTEST} (triangles).  A gradient of
13 MV/m was reached.  The $Q$ was $10^{10}$ at the design gradient of 8 MV/m. 
The field was limited by the available RF power (the input coupling was weaker
than planned).

\section{MICROPHONICS AND MULTIPACTING}

Microphonics are more serious for RIA than for SNS due to the lower RIA beam
current.  The lateral brace and stiffening rings of the SNS $\beta_g = 0.61$
cavity will be used on the $\beta_g = 0.47$ cavity to reduce microphonic
excitation.  The RIA cavities will be over-coupled in order to ensure that the
gradient can be maintained in the presence of microphonics \cite{R_JL}.  Some
microphonic measurements were done on a single-cell cavity \cite{R_JMP}. 
Modelling of the vibrations in multi-cell cavities is in progress.  The
predictions will be compared with measurements on the 6-cell cavity.

The RF tests on single-cell cavities showed that there are no hard multipacting
barriers.  A soft barrier was seen occasionally at very low field. 
Multipacting simulations \cite{R_IIP,R_JNP} also indicate that there should be
no hard barriers in the single-cell cavities.  Likewise, no multipacting
problems were encountered in the tests on the two 6-cell cavities.

\section{CONCLUSION}

RF tests have been done on two single-cell $\beta_g = 0.47$ cavity prototypes
and two 6-cell cavities with encouraging results: all of the cavities exceeded
the desired accelerating gradient, with a $Q \geq 10^{10}$ at the design
gradient of 8 MV/m.  The first 6-cell and both single-cell cavities exceeded
the design gradient by a factor of 2; the second 6-cell reached 13 MV/m.  Two
niobium multi-cells and one copper multi-cell have been tuned for field
flatness.  The next step will be a horizontal test of 2 fully-equipped $\beta_g
= 0.47$ cavities in a prototype cryomodule \cite{R_JQ}.

\section{ACKNOWLEDGEMENTS}

\small

We thank the staff at INFN Milano, Jefferson Lab, and NSCL for their hard work
in the design, fabrication, processing, and testing the cavity prototypes.  
R. Afanador, J. Brawley,  B. Manus, S. Manning, S. Morgan, G. Slack, and L.
Turlington provided essential support with the fabrication and
chemical treatment of the cavities at Jefferson Lab.
J. Bierwagen, J. Brandon, S. Bricker, J. Colthorp, S. Hitchcock, M. Johnson, 
H. Laumer, D. Lawton, A. McCartney, D. Pedtke, L. Saxton,  J. Vincent, and
R. Zink provided valuable support at NSCL.

\end{document}